\documentclass[twocolumn,10pt,letterpaper]{article}
\usepackage{amsopn}
\usepackage{usenix}
\usepackage{graphicx}
\usepackage{caption}
\usepackage[binary,squaren]{SIunits}
\usepackage{tikz}
\usepackage[hyphens]{url}
\usepackage{times}
\usepackage{tabularx}
\usepackage[sort,numbers]{natbib}
\usepackage{enumitem}
\usepackage{amssymb}
\usepackage{listings}
\usepackage[font={small}]{caption}
\usetikzlibrary{calc}
\usetikzlibrary{decorations.pathreplacing}
\usetikzlibrary{patterns}
\usetikzlibrary{shadows}
\usetikzlibrary{arrows}
\usetikzlibrary{positioning}

\newlength\imageheight

\newcommand{\code}[1]{\texttt{#1}}
\newcommand{\myparagraph}[1]{\vspace{-.25\baselineskip}%
\paragraph{#1}}

\usepackage[small,compact]{titlesec}

\newcommand\T{\rule{0pt}{2.6ex}}       
\newcommand\B{\rule[-1.2ex]{0pt}{0pt}} 

\title{The Design and Implementation of the Wave Transactional Filesystem}
\author{\textnormal{Robert Escriva, Emin G\"un Sirer}\\
Computer Science Department, Cornell University
}

\begin{document}

\maketitle

\begin{abstract}
This paper introduces the Wave Transactional Filesystem (WTF), a novel, transactional, POSIX-compatible 
filesystem based on a new {\em file slicing} API that enables efficient file transformations.  
WTF provides transactional access to a distributed filesystem, eliminating the possibility 
of inconsistencies across multiple files. Further, the file slicing API enables applications
to construct files from the contents of other files without having to rewrite or relocate data.
Combined, these enable a new class of high-performance applications.
Experiments show that WTF can qualitatively outperform the industry-standard HDFS distributed filesystem, 
up to a factor of four in a sorting benchmark, by reducing I/O costs. 
Microbenchmarks indicate that the new features of WTF impose only a modest overhead 
on top of the POSIX-compatible API.
\end{abstract}

\section{Introduction}
\label{sec:intro}

Distributed filesystems are a cornerstone of modern data processing
applications.  Key-value stores such as Google's BigTable~\cite{bigtable} and
Spanner~\cite{spanner}, and Apache's HBase~\cite{hbase} use distributed
filesystems for their underlying storage.  MapReduce~\cite{mapreduce} uses a
distributed filesystem to store the inputs, outputs, and intermediary processing
steps for offline processing applications.  Infrastructure such as Amazon's
EBS~\cite{ebs} and Microsoft's Blizzard~\cite{blizzard} use distributed
filesystems to provide storage for virtual machines and cloud-oblivious
applications.

Yet, current distributed filesystems exhibit a tension between retaining the
familiar semantics of local filesystems and achieving high performance in the
distributed setting.  Often, designs will compromise consistency for
performance, require special hardware, or artificially restrict the filesystem
interface.  For example, in GFS, operations can be inconsistent or,
``consistent, but undefined,'' even in the absence of failures~\cite{gfs}.
GFS-backed applications must account for these anomalies, leading to additional
work for application programmers.  HDFS~\cite{hdfs} side-steps this complexity
by prohibiting concurrent or non-sequential modifications to files.  This
obviates the need to worry about nuances in filesystem behavior, but fails to
support use cases requiring concurrency or random-access writes.  Flat
Datacenter Storage~\cite{fds} is only eventually consistent and requires a
network with full-bisection bandwidth, which can be cost prohibitive and is not
possible in all environments.

This paper introduces the Wave Transactional Filesystem (WTF), a new distributed
filesystem that contains a transactional model with a new API that provides {\em
file slicing} operations.  A WTF transaction may span multiple files and is
fully general; applications can include calls such as read, write, and seek
within their transaction.  This file slicing API enables applications to
efficiently read, write, and rearrange files without rewriting the underlying
data.  For example, applications may concatenate multiple files without reading
them; garbage collect and compress a database without writing the data; and even
sort the contents of record-oriented files without rewriting the files'
contents.

The key design decision that enables WTF's advanced feature set is an
architecture that represents filesystem data and metadata to ensure
that filesystem-level transactions may be performed using, solely, transactional
operations on metadata.  Custom storage servers hold filesystem data and handle
the bulk of I/O requests.  These servers retain no information about the
structure of the filesystem; instead, they treat all data as opaque, immutable,
variable-length arrays of bytes, called {\em slices}.  WTF stores references to
these slices in HyperDex~\cite{warp} alongside metadata that describes how to
combine the slices to reconstruct files' contents.  This structure enables most
bookkeeping to be done at the metadata level, within the scope of HyperDex
transactions.

Supporting this architecture is a custom concurrency control layer that
decouples WTF transactions from the underlying HyperDex transactions.  This
layer ensures that applications only abort when a concurrently-executing
transaction changes the filesystem in a way that generates an unresolvable,
application-visible conflict.  This seemingly minor functionality enables WTF to
support many concurrent operations with minimal abort-induced overheads.

Overall, this paper makes three contributions.  First, it describes a new API
for filesystems called file slicing that enables efficient file transformations.
Second, it describes an implementation of a transactional filesystem with
minimal overhead.  Finally, it evaluates WTF and the file slicing interfaces,
and compares them to the non-transactional HDFS filesystem.

\section{Design}

\begin{figure}[t]
\centering
\begin{tikzpicture}
\pgfdeclareimage[height=1cm]{hyperdex}{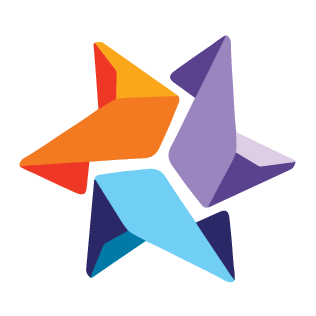};
\settoheight{\imageheight}{\pgfuseimage{hyperdex}};
\pgfdeclareimage[height=1cm]{dbserv}{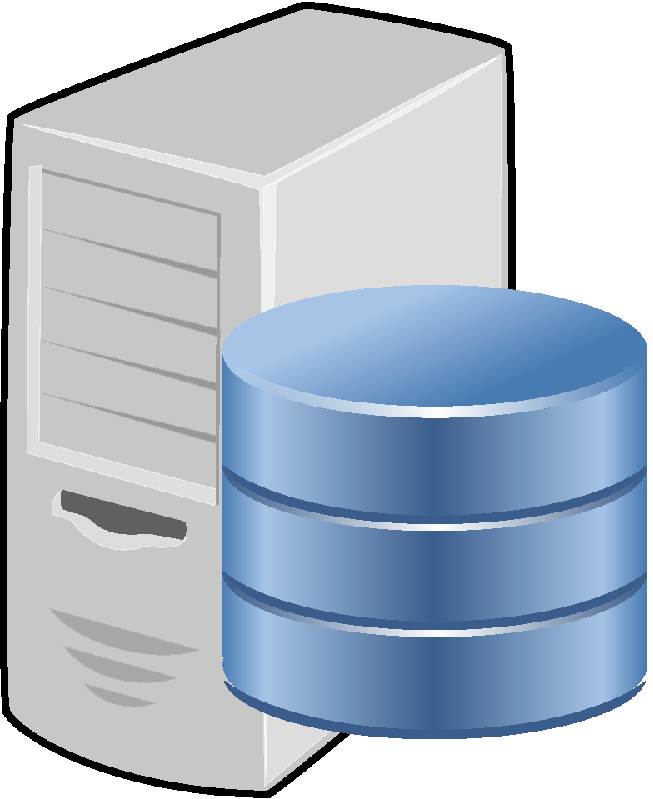};
\settoheight{\imageheight}{\pgfuseimage{dbserv}};
\pgfdeclareimage[height=1cm]{coordserv}{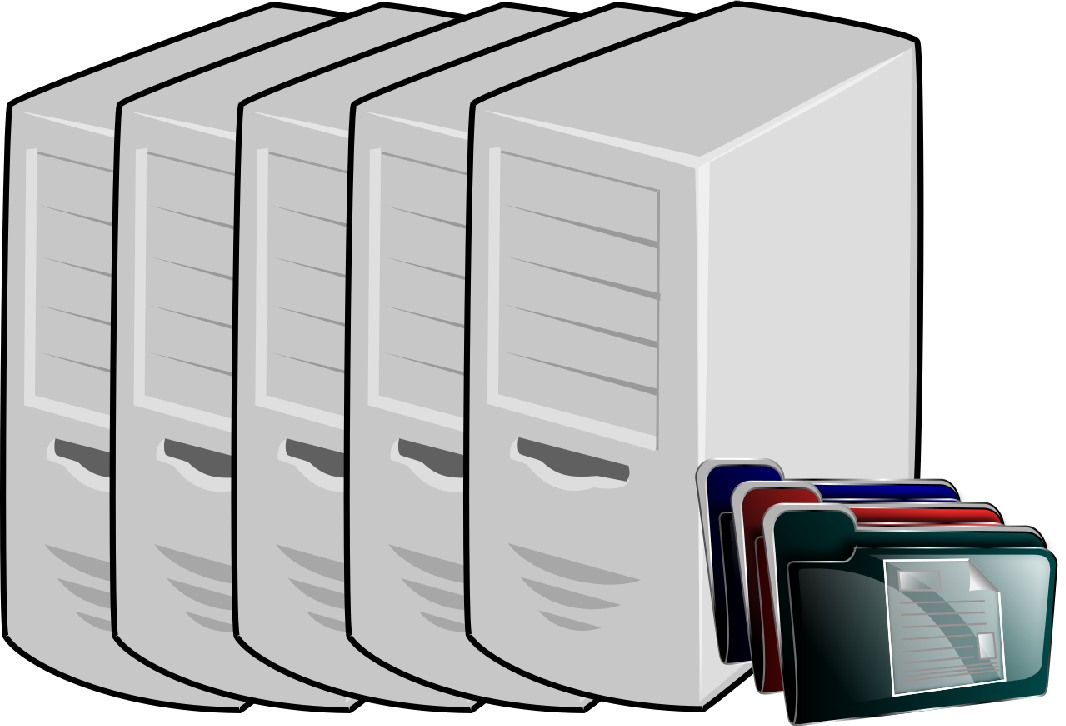};
\settoheight{\imageheight}{\pgfuseimage{coordserv}};
\pgfdeclareimage[height=1cm]{client}{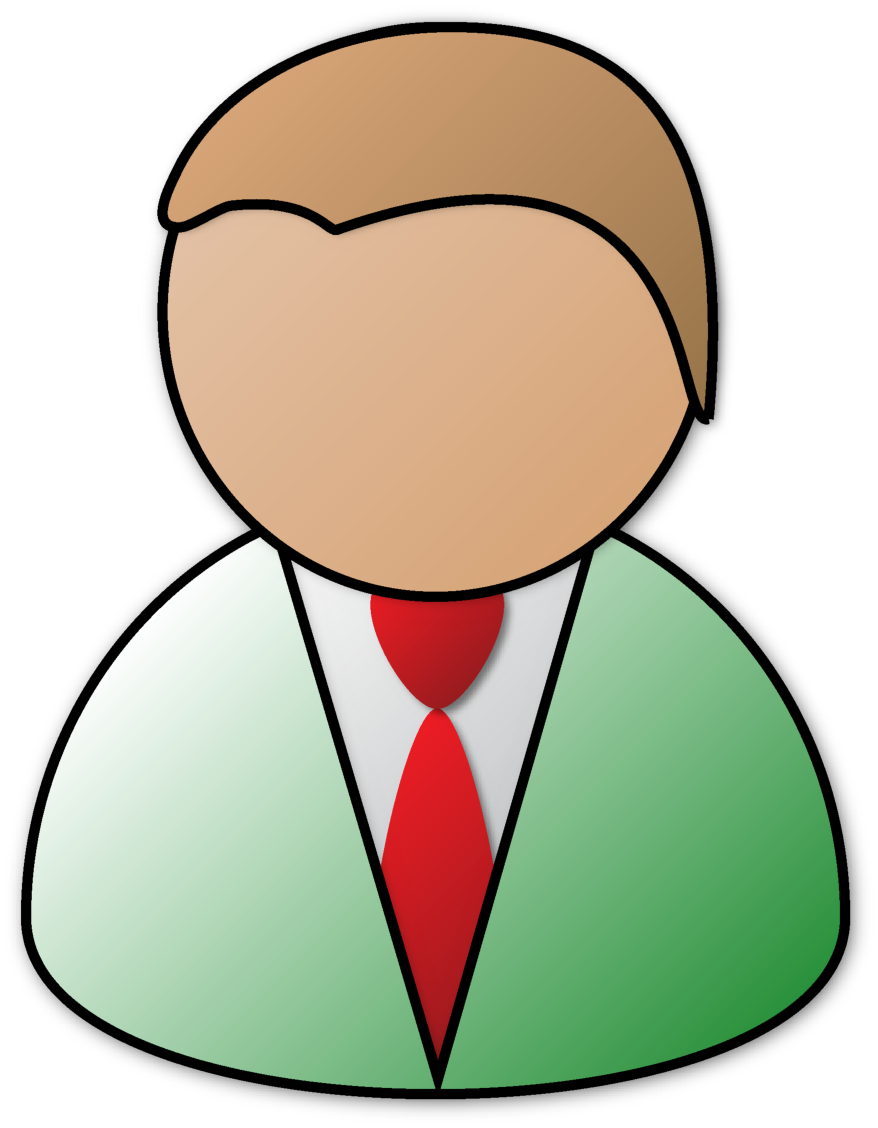};
\settoheight{\imageheight}{\pgfuseimage{client}};
\pgfdeclareimage[height=1cm]{library}{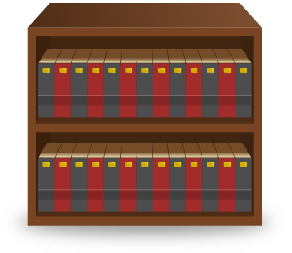};
\settoheight{\imageheight}{\pgfuseimage{library}};

\node[align=center] (hyperdex)
{\pgfuseimage{hyperdex} \\ \normalsize Metadata \\ Storage};

\node[align=center,below of=hyperdex,node distance=7em] (coord)
{\pgfuseimage{coordserv} \\ \normalsize Replicated \\ Coordinator};

\node[align=center,below of=coord,node distance=6em] (storage2) {\pgfuseimage{dbserv}};
\node[align=center,left of=storage2] (storage1) {\pgfuseimage{dbserv}};
\node[align=center,right of=storage2] (storage3) {\pgfuseimage{dbserv}};
\node[align=center,below of=storage2] {\normalsize Storage Servers};

\node[align=center,left of=coord,node distance=8em] (library)
{\pgfuseimage{library} \\ \normalsize Client \\ Library};

\node[align=center,left of=library,node distance=6em] (client)
{\pgfuseimage{client} \\ \normalsize End User \\ Application};

\draw[<->] (client) -- (library);
\draw[<->] (library) -- (hyperdex);
\draw[<->] (library) -- (coord);
\draw[<->] (library) -- (storage1);
\draw[<->] (storage1) -- (coord);
\draw[<->] (storage2) -- (coord);
\draw[<->] (storage3) -- (coord);
\draw[<->] (hyperdex) -- (coord);

\end{tikzpicture}
\vspace{-.75\baselineskip}
\caption{WTF employs a distributed architecture consisting of metadata storage,
data storage, a replicated coordinator, and the client library.}
\label{fig:arch}
\vspace{-\baselineskip}
\end{figure}
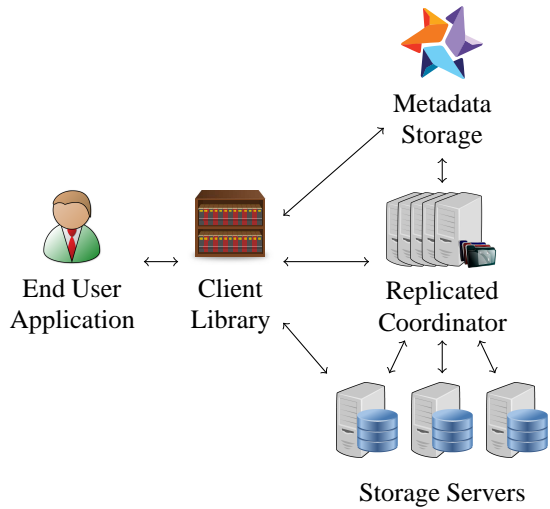

WTF's distributed architecture consists of four components:  the metadata
storage, the storage servers, the replicated coordinator, and the client
library.  Figure~\ref{fig:arch} summarizes this architecture.  The metadata
storage builds on top of HyperDex and its expansive API.  The storage servers
hold filesystem data, and are provisioned for high I/O workloads.  A replicated
coordinator service serves as a rendezvous point for all components of the
system, and maintains the list of storage servers.  The client library contains
the majority of the functionality of the system, and is where WTF combines the
metadata and data into a coherent filesystem.

In this section, we first explore the file slicing abstraction to understand how
the different components contribute to the overall design.  We will then look at
the design of the storage servers to understand how the system stores the
majority of the filesystem information.  Finally, we discuss performance
optimizations and additional functionality that make WTF practical, but are not
essential to the core design, such as replication, fault tolerance, and garbage
collection.

\subsection{The File Slicing Abstraction}

WTF represents a file as a sequence of byte arrays that, when overlaid, comprise
the file's contents.  The central abstraction is a {\em slice}, an immutable,
byte-addressable, arbitrarily sized sequence of bytes.  A file in WTF, then is a
sequence of slices and their associated offsets.  This representation has some
inherent advantages over block-based designs.  Specifically, the abstraction
provides a separation between metadata and data that enables filesystem-level
transactions to be implemented using, solely, transactions over the metadata.
Data is stored in the slices, while the metadata is a sequence of slices.  WTF
can transactionally change these sequences to change the files they represent,
without having to rewrite the data.

Concretely, file metadata consists of a list of {\em slice pointers} that
indicate the exact location on the storage servers of each slice.  A slice
pointer is a tuple consisting of the unique identifier for the storage server
holding the slice, the local filename containing the slice on that storage
server, the offset of the slice within the file, and the length of the slice.
Associated with each slice pointer is an integer offset that indicates where the
slice should be overlaid when reconstructing the file.  Crucially, this
representation is self-contained: everything necessary to retrieve the slice
from the storage server is present in the slice pointer, with no need for extra
bookkeeping elsewhere in the system.  As we will discuss later, the metadata
also contains standard info found in an inode, such as modification time, and
file length.

This slice pointer representation enables WTF to easily generate new slice
pointers that refer to subsequences of existing slices.  Because the
representation transparently reflects the global location of a slice on disk,
WTF may use simple arithmetic to create new slice pointers.

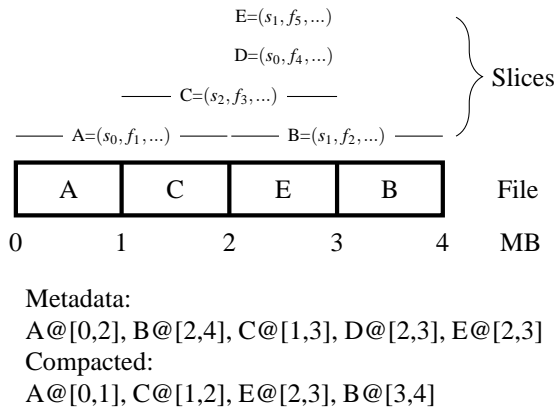
\begin{figure}[t]
\centering
\begin{tikzpicture}

\node (one) {};
\node[right of=one,node distance=4em] (two) {};
\node[right of=two,node distance=4em] (three) {};
\node[right of=three,node distance=.075em] (threehalf) {};
\node[right of=three,node distance=.15em] (four) {};
\node[right of=four,node distance=2em] (fourhalf) {};
\node[right of=four,node distance=4em] (five) {};
\node[right of=five,node distance=4em] (six) {};

\draw[-,thin] ($(one)+(0,2em)$) -- ($(three)+(0,2em)$) node[midway,fill=white] {\scriptsize A=$(s_0, f_1, ...)$};
\draw[-,thin] ($(four)+(0,2em)$) -- ($(six)+(0,2em)$) node[midway,fill=white] {\scriptsize B=$(s_1, f_2, ...)$};
\draw[-,thin] ($(two)+(0,3.5em)$) -- ($(five)+(0,3.5em)$) node[midway,fill=white] {\scriptsize C=$(s_2, f_3, ...)$};
\draw[-,thin] ($(four)+(0,5em)$) -- ($(five)+(0,5em)$) node[midway,fill=white] {\scriptsize D=$(s_0, f_4, ...)$};
\draw[-,thin] ($(four)+(0,6.5em)$) -- ($(five)+(0,6.5em)$) node[midway,fill=white] {\scriptsize E=$(s_1, f_5, ...)$};

\draw[ultra thick] ($(one) + (0, 1em)$) rectangle ($(six) + (0, -1em)$);
\node[right of=six] (file) {File};

\node[node distance=2em,below of=one] {0};
\node[node distance=2em,below of=two] {1};
\node[node distance=2em,below of=threehalf] {2};
\node[node distance=2em,below of=five] {3};
\node[node distance=2em,below of=six] {4};
\node[node distance=2em,below of=file] {\unit{}{\mega\byte}};

\draw[ultra thick] ($(two) + (0, 1em)$) -- ($(two) + (0, -1em)$);
\draw[ultra thick] ($(threehalf) + (0, 1em)$) -- ($(threehalf) + (0, -1em)$);
\draw[ultra thick] ($(five) + (0, 1em)$) -- ($(five) + (0, -1em)$);

\draw[color=white,text=black] (one) -- (two) node[midway] {A};
\draw[color=white,text=black] (two) -- (threehalf) node[midway] {C};
\draw[color=white,text=black] (threehalf) -- (five) node[midway] {E};
\draw[color=white,text=black] (five) -- (six) node[midway] {B};

\draw [decorate,decoration={brace,mirror,amplitude=10pt}] ($(six) + (.5em,2em)$) --
($(six) + (.5em,6.5em)$) node[align=left,right,midway,xshift=10pt] {Slices};

\node[anchor=north,align=left,below of=fourhalf,node distance=6em]
{Metadata: \\
A@[0,2], B@[2,4], C@[1,3], D@[2,3], E@[2,3] \\
Compacted: \\
A@[0,1], C@[1,2], E@[2,3], B@[3,4]};

\end{tikzpicture}
\caption{A \unit{4}{\mega\byte} file with five writes that write or overwrite
different portions of the file.  This figure shows the slices that were written,
the resulting file's content, and the metadata in HyperDex.}
\label{fig:file-slicing}
\vspace{-.5\baselineskip}
\end{figure}

This representation also enables applications to modify a file with only
localized modifications to the metadata.  Figure~\ref{fig:file-slicing} shows an
example file consisting of five different slices.  Each slice is overlaid on top
of previous slices.  Where slices overlap, the latest additions to the metadata
take precedence.  For example, slice $C$ takes precedence over slices $A$ and
$B$; similarly, slice $E$ completely obscures slice $D$ and part of $C$.  The
file, then, consists of the corresponding slices of $A$, $C$, $E$, and $B$.  The
figure also shows the {\em compacted} metadata for the same file.  This
compacted form contains the minimal slice pointers necessary to reconstruct the
file without reading data that is overwritten by another slice.  Crucially, file
modifications can be performed without having to rearrange the entire metadata.

The procedures for reading and writing follow directly from the abstraction.  A
writer creates one or more slices on the storage servers, and overlays them at
the appropriate positions within the file by appending their slice pointers to
the metadata list.  Readers retrieve the metadata list, compact it, and
determine which slices must be retrieved from the storage servers to fulfill the
read.

The correctness of this design relies upon the metadata storage providing
primitives to atomically read and append to the list.  HyperDex natively
supports both of these operations.  Because each writer writes slices before
appending to the metadata list, it is guaranteed that any transaction that can
see these immutable slices is serialized {\em after} the writing transaction
commits.  It can then retrieve the slices directly.  The transactional
guarantees of WTF extend directly from this design as well:  a WTF transaction
will execute a single HyperDex transaction consisting of multiple append and
retrieve operations.

\subsection{Storage Server Interface}

The file slicing abstraction greatly simplifies the design of the storage
servers.  Storage servers deal exclusively with slices, and are oblivious to
files, offsets, or concurrent writes.  Instead, the complete storage server API
consists of just two calls that create and retrieve slices.

A storage server processes a request to create a slice by writing the data to
disk and returning a slice pointer to the caller.  The structure of this request
intentionally grants the storage server complete flexibility to store the slice
anywhere it chooses because the slice pointer containing the slice's location is
returned to the client only after the slice is written to disk.  A storage
server can retrieve slices by following the information in the slice pointer to
open the named file, read the requisite number of bytes, and return them to the
caller.

The transparency of the slice pointer minimizes the bookkeeping of the storage
server implementation, while also permitting a wide variety of implementation
strategies.  Currently, each WTF storage server maintains a directory of
slice-containing backing files and information about their own identities in the
system.  Each backing file is written sequentially as the storage server creates
new slices.

As an optimization, the storage servers maintain multiple backing files to which
slices are written.  This serves three purposes:  First, it allows servers to
avoid contention when writing to the same file; second, it allows the storage
server to explicitly spread data across multiple filesystems if configured to do
so; and, finally, it allows the storage server to use hints provided by writers
to improve locality within backing files, as described in
Section~\ref{sec:locality}.

\subsection{File Partitioning}

Practically, it is desirable to keep the list of slice pointers small so that they
can be stored, retrieved, and transmitted with low overhead; however, it would
be impractical to achieve this by limiting the number of writes to a file.  In
order to achieve support for both arbitrarily large files and efficient
operations on the list of slice pointers, WTF partitions a file into fixed size
regions, each with its own list.  Each region is stored as its own object in
HyperDex under a deterministically derived key.

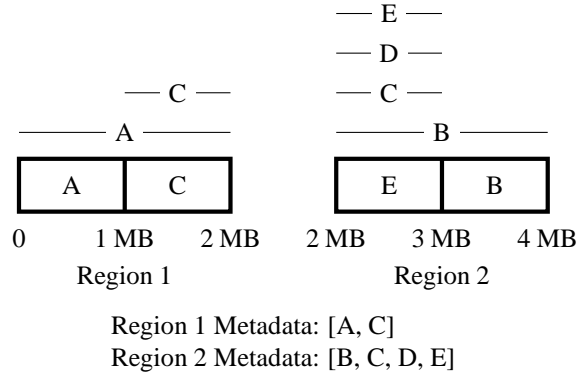
\begin{figure}[t]
\centering
\begin{tikzpicture}

\node (one) {};
\node[right of=one,node distance=4em] (two) {};
\node[right of=two,node distance=4em] (three) {};
\node[right of=three,node distance=2em] (sep) {};
\node[right of=sep,node distance=2em] (four) {};
\node[right of=four,node distance=4em] (five) {};
\node[right of=five,node distance=4em] (six) {};

\draw[ultra thick] ($(one) + (0, 1em)$) rectangle ($(three) + (0, -1em)$);
\draw[ultra thick] ($(four) + (0, 1em)$) rectangle ($(six) + (0, -1em)$);
\draw[ultra thick] ($(two) + (0, 1em)$) -- ($(two) + (0, -1em)$);
\draw[ultra thick] ($(five) + (0, 1em)$) -- ($(five) + (0, -1em)$);

\node[below of=two,node distance=3.5em] (seg1) {\normalsize Region 1};
\node[below of=five,node distance=3.5em] (seg2) {\normalsize Region 2};

\draw[-,thin] ($(one)+(0,2em)$) -- ($(three)+(0,2em)$) node[midway,fill=white] {A};
\draw[-,thin] ($(four)+(0,2em)$) -- ($(six)+(0,2em)$) node[midway,fill=white] {B};
\draw[-,thin] ($(two)+(0,3.5em)$) -- ($(three)+(0,3.5em)$) node[midway,fill=white] {C};
\draw[-,thin] ($(four)+(0,3.5em)$) -- ($(five)+(0,3.5em)$) node[midway,fill=white] {C};
\draw[-,thin] ($(four)+(0,5em)$) -- ($(five)+(0,5em)$) node[midway,fill=white] {D};
\draw[-,thin] ($(four)+(0,6.5em)$) -- ($(five)+(0,6.5em)$) node[midway,fill=white] {E};

\draw[color=white,text=black] (one) -- (two) node[midway] {A};
\draw[color=white,text=black] (two) -- (three) node[midway] {C};
\draw[color=white,text=black] (four) -- (five) node[midway] {E};
\draw[color=white,text=black] (five) -- (six) node[midway] {B};

\node[node distance=2em,below of=one] {0};
\node[node distance=2em,below of=two] {\unit{1}{\mega\byte}};
\node[node distance=2em,below of=three] {\unit{2}{\mega\byte}};
\node[node distance=2em,below of=four] {\unit{2}{\mega\byte}};
\node[node distance=2em,below of=five] {\unit{3}{\mega\byte}};
\node[node distance=2em,below of=six] {\unit{4}{\mega\byte}};

\node[align=left,below of=sep,node distance=6em]
{\normalsize Region 1 Metadata: [A, C] \\ Region 2 Metadata: [B, C, D, E]};

\end{tikzpicture}
\caption{A file in WTF that is partitioned into \unit{2}{\mega\byte} regions.
Writes within each region are appended solely to that region's metadata.
Writes that cross regions, like $C$, are atomically applied to both lists.}
\label{fig:partitioning}
\vspace{-.5\baselineskip}
\end{figure}

\begin{table*}[t]
\centering
\begin{tabular}{ll}
\hline
\T\B API & Description \\
\hline
\T \code{yank(fd,sz):slice,[data]} & Copy \code{sz} bytes from \code{fd}; return
slice pointers and optionally the data \\
\code{paste(fd, slice)} & Write \code{slice} to \code{fd} and increment the
offset\\
\code{punch(fd, amount)} & Zero-out \code{amount} bytes at the fd offset, freeing the underlying storage\\
\B \code{append(fd, slice)} & Append \code{slice} to the end of file \code{fd} \\
\hline
\T \code{concat(sources, dest)} & Concatenate the listed files to create dest \\
\B \code{copy(source, dest)} & Copy source to dest using only the metadata \\
\hline
\end{tabular}
\vspace{-.5\baselineskip}
\caption{WTF's new file slicing API.  Note that these supplement the POSIX API,
which includes calls for moving a file descriptor's offset via \code{seek}.
\code{concat} and \code{copy} are provided for convenience and may be
implemented with \code{yank} and \code{paste}.}
\label{tab:apis}
\vspace{-\baselineskip}
\end{table*}

Operations on these partitioned metadata lists behave the same as operations on
a single list.  When operations span multiple regions, they are separated into
their respective operations on each region, and performed within the context of
a single multi-key HyperDex transaction.  This guarantees that multiple regions
may be modified simultaneously in one atomic action.
Figure~\ref{fig:partitioning} shows a series of writes that span different
metadata regions, and their resulting metadata lists.

\subsection{Filesystem Hierarchy}

The WTF filesystem hierarchy is modeled after the traditional Unix filesystem,
with directories and files.  Each directory contains entries that are
named links to other directories or files, and WTF enables files to be 
hard linked to multiple places in the filesystem hierarchy.

WTF implements a few changes to the traditional filesystem behavior to reduce
false dependencies when opening a file.  If one were to implement path traversal
as it is traditionally implemented, an open operation would require a traversal
from the root, putting every directory along the path within the scope of a
transaction, and require several round trips to both HyperDex and the storage
servers to open a file.

WTF avoids traversing the filesystem on open by maintaining a pathname to inode
mapping.  This enables a client to map a pathname to the corresponding inode
with just one HyperDex lookup, no matter how deeply nested the pathname.  To
enable applications to enumerate the contents of a single directory, WTF
maintains traditional-style directories, implemented as special files, alongside
the one-lookup mapping.  The two data structures are atomically updated using
HyperDex transactions.  This optimization simplifies the process of opening
files, without a loss of functionality.

Inodes are also stored in HyperDex, and contain standard information, such as
link count and modification time.  The inode also maintains ownership, group,
and permissions information, though WTF differs from POSIX in that permissions
are not checked on the full pathname from the root.  Each inode also stores a
reference to the highest-offset region written within the file, enabling
applications to find the end of the file.

Because HyperDex permits transactions to span multiple keys across independent
schemas, updates to the filesystem hierarchy remain consistent.  For example, to
create a hardlink for a file, WTF atomically creates a new pathname to inode
mapping for the file, increments the inode's link count, and inserts the
pathname and inode pair into the destination directory, which requires a write
to the file holding the directory entries.

\subsection{File Slicing Interface}

The file slicing interface enables new applications to make more efficient use
of the filesystem.  Instead of operating on bytes and offsets as traditional
POSIX systems do, this new API allows applications to manipulate subsequences of
files at the structural level, without copying or reading the data itself.

Table~\ref{tab:apis} summarizes the new APIs that WTF provides to applications.
The \code{yank}, \code{paste}, and \code{append} calls are analogous to read,
write, and append, but operate on slices instead of sequences of bytes.  The
\code{yank} call retrieves slice pointers for a range of the file.  An
application may provide these slice pointers to a subsequent call to
\code{paste} or \code{append} to write the data back to the filesystem, reusing
the existing slices.  These write operations bypass the storage servers and only
incur costs at the metadata storage component.

The \code{append} call is internally optimized to improve throughput.  A naive
\code{append} call could be implemented as a transaction that seeks to the end
of the file, and performs a \code{paste}.  While not incorrect, it would allow
only one append call to proceed at a time, because only one append can commit
for each value for the end of file; the others will spuriously fail and retry.
Instead, WTF stores, alongside the metadata list, an offset representing the end
of the region.  An \code{append} call will conditionally append to the list,
making sure that the offset, plus the length of the slice to be appended, does
not exceed the bounds of the metadata region.  The entry in the metadata list
for an append is marked as relative to the end of the file, rather than a
specific offset.  When an append is too large to fit within a single region, WTF
will fall back on reading the offset of the end of file, and performing a write
at that offset.  This enables multiple \code{append} operations to proceed in
parallel in the common case.

Other calls that are new to the file slicing API have no counter-part in
traditional APIs.  The \code{concat} call concatenates multiple files to create
one unified output file.  The \code{copy} call creates a copy of a file by
copying the file's compacted metadata.  Both of these calls may be implemented
by \code{yank} and \code{paste} and are provided for convenience.

\subsection{Transaction Retry}

To ensure that transactions abort only when they encounter application-visible
conflicts, WTF implements its own concurrency control on top of HyperDex that
retries aborted transactions.  To see why this may be necessary, consider an
application that seeks to the end of a file, and writes the string ``Hello
World'' within a single transaction.  Barring any permanent failures, such a
transaction should always succeed because this transaction can serialize between
any other pair of transactions as it does not impose any requirements on the
filesystem state.  If, however, a write were to change the length of the file
between the end-of-file lookup and the transaction commit, the transaction
encompassing the original seek-and-write operation will abort within HyperDex
because the observed value of the file length has changed.  Passing this failure
up to the application, which never saw the offset of the end of file, would
complicate the guarantees made by the WTF interface.  Instead, WTF internally
retries the transaction by repeating the seek and then pasting the previously
written slice that contains ``Hello World'' at the new end of file.  This
ensures that transactions only abort in response to an unresolvable,
application-visible conflict.

The mechanism that retries transactions is a thin layer that sits at the
boundary of the WTF client library and the user's application.  Each call the
application makes is logged, along with the arguments provided to the call, and
its return value.  If the transaction aborts within HyperDex, the state of the
system remains unchanged by the WTF transaction, so it is safe to retry it in
its entirety.  WTF will then replay all of the user's operations in sequence
using the same arguments originally supplied.  If at any point a re-executed
call completes with an outcome different from the original execution, the
transaction will signal an abort to the application.  Similarly, if the WTF
transaction re-executes all operations successfully, and the HyperDex
transaction commits, the commit status is passed back to the application.  WTF
will retry transactions as necessary to ensure that they only abort when
operations on the filesystem generate unresolvable, application-visible
conflicts.

To reduce the overhead for maintaining the log of individual operations, the
client library uses slice pointers to refer to bytes of data that pass through
the interface.  For example, a write of \unit{100}{\mega\byte} will not be
copied and maintained in the log; instead, the log maintains the slice pointers
that refer to the \unit{100}{\mega\byte} on the storage servers.  Similarly,
reads are maintained using the retrieved slice pointers, and not the data itself
or checksums thereof.

\subsection{Locality-Aware Slice Placement}
\label{sec:locality}

As an optimization, WTF employs a locality-aware slice placement algorithm to
improve the locality on disk of writes to nearby ranges of a file.  Writes to
the same metadata region reside on the same servers, and are located near each
other on those servers' disks.  Files that are written to WTF sequentially will,
with high probability, be written sequentially to disk.

WTF chooses which server to write a slice to using consistent
hashing~\cite{consist-hash} across the servers to ensure that writes to the same
region reside on the same storage server.  The writer provides the slice and the
identity of the metadata region the write affects to the storage server, which
then uses consistent hashing to map each slice to a file on its local disk.  The
hashing function used at the storage server level is different from the hashing
function used across storage servers, so writes which map to the same server
will be unlikely to map to the same backing file, unless they are for the same
metadata region.

Overall, this ensures that a writer that writes sequentially to a file will
write contiguous sequences of bytes on the storage servers with high
probability.  During compaction, these independent slices may be combined into a
single slice spanning the maximum contiguous range on the disk.  For example, a
sequential writer writing fixed size \unit{1}{\mega\byte} blocks to a metadata
region will sequentially send each of these blocks to the same storage server,
which will append them to the same file on disk.  These adjacent slices may be
compactly represented by a single slice pointer that references the contiguous
region.

\subsection{Garbage Collection}
\label{sec:gc}

WTF prevents unbounded growth of data and metadata through a three-tiered garbage collection
mechanism.  

First, the most prevalent form of garbage in WTF comes from the metadata lists growing
when many independent append operations force it to grow.  This predominant case
is easily handled by compacting the metadata list, and storing the compacted
list in place of the original list.  This eliminates the garbage generated from
overlaid slices, such as those in Figure~\ref{fig:file-slicing}, and will
typically combine multiple slices into one because of locality-aware slice placement.  WTF
retrieves the current metadata list, compacts it, and stores the result using a
single HyperDex transaction.  The resulting file contents are equivalent to
those from before the compaction, and the compaction incurs no I/O on the
storage servers.

Metadata compaction is not always sufficient. In particular,
random writes reduce the effect that locality-aware slice placement has on
compaction, leading to fragmented metadata lists.  In this case, WTF writes a
new slice with contents identical to the compacted form of the current metadata
list, and swaps a pointer to this slice with the originally observed list.

Finally, as an application overwrites or deletes files, slices become unused by the
filesystem and turn into garbage on the storage servers.  Because the storage
servers outsource all bookkeeping to the metadata storage, storage servers do
not directly know which portions of its local data are garbage.  WTF
periodically scans the entire filesystem metadata and constructs a list of
in-use slice pointers for each storage server.  For simplicity of
implementation, these lists are stored in a reserved directory within the WTF
filesystem so that they need not be maintained in memory or communicated out of
band to the storage servers.  Storage servers link the WTF client library and
read their respective files to discover unused regions in their local storage
space.  To prevent the race condition where a slice is created and garbage
collected before being referenced by the metadata, the periodic garbage
collection is run infrequently---on the order of hours or days---and servers do
not collect an unused region until it appears in two consecutive scans.

Storage servers implement garbage collection by creating sparse files on the
local disk.  To compress a file containing garbage slices, a storage server
rewrites the file, seeking past each unused slice.  On inode-based Linux
filesystems this creates a sparse file that occupies disk space proportional to
the in-use slices it contains.  Counter-intuitively, files with the most garbage
are the most efficient to collect, because the garbage collection thread seeks
past large regions of garbage and only writes the small number of remaining
slices.  Backing files with little garbage incur much more I/O, because there are more
in-use slices to rewrite.  WTF chooses the file with the most garbage to compact
first, because it will simultaneously compact the most garbage and incur the
least I/O. 

The storage servers derive benefit from the kernel buffer cache by relying upon
writing to a local filesystem rather than direct disk access. When writing a
file, Linux will not start to flush the data to disk immediately, but will
instead flush data in batched writes.  The filesystem coalesces many writes and
reduces the number of seeks used by garbage collection~\cite{ext4}.

\subsection{Fault Tolerance}

WTF uses replication to add a configurable degree of fault tolerance to the
system.  To accomplish this, it augments the metadata list such that each entry
references multiple slice pointers that are replicas of the data.  On the write
path, writers create multiple replica slices and append their pointers
atomically.  Readers may read from any of the replicas, as they hold identical
data.

The metadata storage derives its fault tolerance from the strong guarantees
offered by HyperDex.  Specifically, HyperDex guarantees that it can tolerate $f$
failures for a user-configurable value of $f$.  HyperDex uses value-dependent
chaining to coordinate between the replicas and manage recovery from
failures~\cite{hyperdex}.

The data storage derives its durability guarantees from the backing file system.
While replication protects WTF against uncorrelated failures, WTF is not designed 
to withstand correlated failures such as cluster-wide power outages.

The file slicing abstraction is easier to make fault tolerant and consistent
than existing block-based solutions.  In a block-based design a write is often
constrained to reuse existing replicas for the block it is writing.  Further,
block designs often employ some mechanism on top of the block servers to
consistently update all replicas, or at least ensure they eventually converge to
the same value.  This added mechanism introduces overheads that are absent in
WTF's slice-based design.

\section{Implementation}

Everything described in this paper is available in our WTF implementation.
Currently, the implementation is approximately \unit{30}{\kilo} lines of code
written exclusively for WTF.  It relies upon HyperDex with transactions, which
is approximately \unit{85}{\kilo} lines of code, with an additional
\unit{37}{\kilo} lines of code of supporting libraries written for both
projects.  The replicated coordinator for both HyperDex and WTF is an additional
\unit{19}{\kilo} lines of code.  Altogether, WTF constitutes \unit{171}{\kilo}
lines of code that were written for WTF and HyperDex.

WTF's fault tolerant coordinator is implemented as a replicated object on top of
the Replicant replicated state machine service.  The coordinator consists of
just 960 lines of code that are compiled into a dynamically linked library that
is passed to Replicant.  Replicant deploys multiple copies of the library, and
uses Paxos~\cite{paxos} to sequence the function calls into the library.

\section{Evaluation}
\label{sec:eval}

To evaluate WTF, we will look at a series of both end-to-end and micro
benchmarks that demonstrate our working implementation under a variety of
conditions.  The first part of this section looks at the how the file slicing
interface improves an end-to-end sorting benchmark written in the style of a map
reduce application.  We will then look at a series of microbenchmarks that
characterize the performance of WTF's conventional filesystem interface.

All benchmarks execute on a cluster of fifteen servers dedicated to the
experiment.  Each server is equipped with two Intel Xeon \unit{2.5}{\giga\hertz}
L5420 processors, \unit{16}{\giga\byte} of DDR2 memory with error correction,
and between \unit{500}{\giga\byte} and \unit{1}{\tera\byte} SATA spinning-disks
from the same era as the CPUs.  The servers are connected with gigabit ethernet
via a single top of rack switch.  Installed on each server is 64-bit Ubuntu
14.04, HDFS from Apache Hadoop 2.7, and WTF with HyperDex 1.8.1.

For all benchmarks, HDFS and WTF are configured to provide an apples-to-apples
comparison.  Both systems are deployed with three nodes reserved for the
meta-data---the HDFS name node, or the HyperDex cluster---and the remaining
twelve servers are allocated as storage nodes for the data.  Except for changes
necessary to achieve feature parity, both systems were deployed in their default
configuration.  To bring the semantics of HDFS up to par with WTF, each
\code{write} is followed by an \code{hflush} call to ensure that the write is
flushed from the client-side buffer and is visible to readers.  The
\code{hflush} primitive solely makes sure that writes are visible to all
readers, and does {\em not} trigger an \code{fsync} on the written data; the
resulting guarantee is the same guarantee provided by a WTF write, and no
stronger.

Additionally, in order to work around a long-standing bug with append
operations~\cite{hdfs-append-bug}, the HDFS block size was reduced from
\unit{128}{\mega\byte} to \unit{64}{\mega\byte}.  Without this change to the
configuration, the HDFS node can report an out-of-disk-space condition when only
3\% of the disk space is in use.  Instead of gracefully handling the condition
and falling back to other replicas as is done in WTF, the failure cascades and
causes multiple writes to fail, making it impossible to complete the benchmark.
Decreasing the block size does increase the amount of metadata held on the name
node, but because all data is held within main memory, and our workloads do not
generate more metadata than the HDFS name node's memory capacity, the increase
is irrelevant to our benchmarks.  The change is unlikely to impact the
performance of data nodes because the increase from
\unit{64}{\mega\byte} to \unit{128}{\mega\byte} was not motivated by
performance~\cite{hdfs-block-size}.  WTF is also configured to
use \unit{64}{\mega\byte} regions.

\begin{table}
\centering
\begin{tabular}{lll}
\hline
\T\B Stage & Conventional & File Slicing \\
\hline
\T Bucketing & R = \unit{100}{\giga\byte} & R = \unit{100}{\giga\byte} \\
          & W = \unit{100}{\giga\byte} & W = \unit{0}{\giga\byte} \\
Sorting   & R = \unit{100}{\giga\byte} & R = \unit{100}{\giga\byte} \\
          & W = \unit{100}{\giga\byte} & W = \unit{0}{\giga\byte} \\
Merging   & R = \unit{100}{\giga\byte} & R = \unit{0}{\giga\byte} \\
\B        & W = \unit{100}{\giga\byte} & W = \unit{0}{\giga\byte} \\
\hline
\T Total  & R = \unit{300}{\giga\byte} & R = \unit{200}{\giga\byte} \\
\B        & W = \unit{300}{\giga\byte} & W = \unit{0}{\giga\byte} \\
\hline
\end{tabular}
\vspace{-.5\baselineskip}
\caption{File slicing enables the WTF-based sort application to sort a
\unit{100}{\giga\byte} file with one third the I/O required by conventional
distributed filesystems.}
\label{tab:sort:io}
\vspace{-\baselineskip}
\end{table}

Except where otherwise noted, both systems replicate all files such that two
copies of the file exist.  This allows the filesystem to tolerate the failure of
any one storage server throughout the experiment without loss of data or
availability.  It is possible to tolerate more failures so long as all the
replicas for a file do not fail simultaneously.

\subsection{Map Reduce: Sorting}

MapReduce~\cite{mapreduce} is a processing technique that forms the basis of
many modern analytic applications.  Because filesystems like HDFS and GFS are
the basis of modern mapreduce frameworks, mapreduce applications provide a
useful means of evaluating new distributed filesystems.

Sorting a file with mapreduce is a three-step process that breaks the sort into
two map jobs followed by a reduce job.  The first map task partitions the input
file into buckets, each of which holds a disjoint, contiguous section of the
keyspace.  These buckets are sorted in parallel by the second map task.
Finally, the reduce phase concatenates the sorted buckets to produce the sorted
output.

Each intermediate step of this application is written to disk, implying that the
entire data set will be read or written several times over.  Here, WTF's file
slicing interface can reduce this excessive I/O and improve the efficiency of
the application.  Instead of reading and writing whole records during the first
two stages, WTF can use \code{yank} and \code{paste} to rearrange the records.
File slicing also eliminates almost all I/O of the reduce phase using a
\code{concat} operation.  Table~\ref{tab:sort:io} summarizes the number of bytes
of data we can expect to be read or written while sorting a
\unit{100}{\giga\byte} file.  We can see that a conventional API will perform
\unit{600}{\giga\byte} of total I/O while a file-slicing filesystem can do the
same task with only \unit{200}{\giga\byte} of I/O.

\begin{figure}[t]
\centering
\input{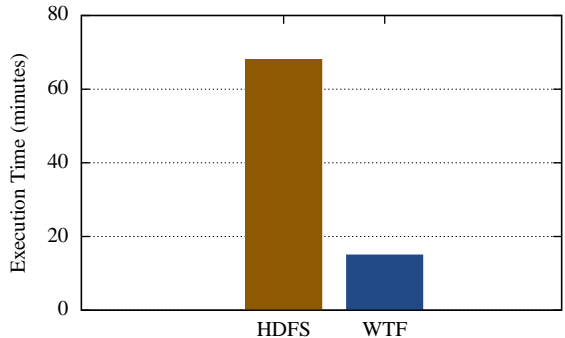}
\caption{Total execution time for sorting a \unit{100}{\giga\byte} file using a
    map-reduce application.  HDFS takes more than one hour and seven minutes to
    sort the file, while WTF completes the same task in under fifteen minutes.}
\label{fig:sort:summary}
\vspace{-.5\baselineskip}
\end{figure}

Empirically, the file slicing operations do improve the running time of a
WTF-based sort.  Figure~\ref{fig:sort:summary} shows the total running time of
both systems to sort a \unit{100}{\giga\byte} file consisting of
\unit{500}{\kilo\byte} records indexed by \unit{10}{\byte} keys that were
generated uniformly at random.  In this benchmark, the intermediate files are
written without replication because they may easily be recomputed from the
input.  We can see that WTF sorts the entire file in one fourth the time taken
to perform the same task on HDFS.

\begin{figure}[t]
\centering
\input{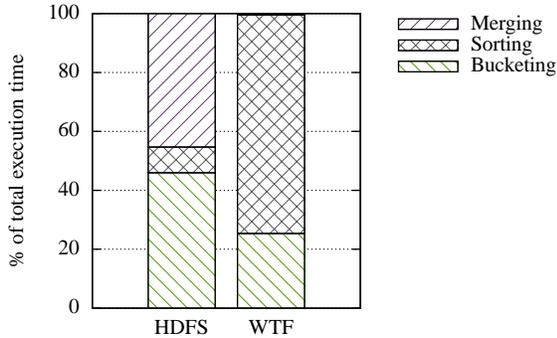}
\caption{Execution time of the sort broken down by stage of the map-reduce
application.  HDFS spends 91.5\% of its time partitioning and reassembling
the data, compared to WTF, which spends 25.9\% of its time on the same task.}
\label{fig:sort:alloc}
\vspace{-\baselineskip}
\end{figure}

The speedup is attributable to the efficient primitives that WTF exposes to
applications.  From Figure~\ref{fig:sort:alloc}, we can see that the WTF-based
sorting application spends less time in the partitioning and merging steps than
the conventional HDFS-based application.  For HDFS, the majority of execution
time is spent in merging and bucketing of the input data.  Just 8.5\% of
execution time is spent in the CPU-intensive sorting task.  The rest is spent
shuffling data on either side of this task.  In contrast, WTF spends 74.1\% of
its time in the CPU intensive task, whereas the first map phase accounts for
25.3\% of the execution time.  The concatenation operation at the end occupies
less than 1\% of the overall running time.  From this, we can conclude that the
efficiency of WTF's I/O operations contribute to reducing the overall runtime of
the sort operation.

Overall this sorting benchmark shows that file slicing operations can improve
map reduce performance.  In general, applications that process data by
partitioning, shuffling, or combining records will benefit from a reduction in
I/O and decrease in running time.

\subsection{Micro Benchmarks}

In this section we examine a series of microbenchmarks that quantify the
performance of the POSIX API for both HDFS and WTF.  Here HDFS serves as a
gold-standard.  With ten years of active development, and deployment across
hundreds of nodes, including large deployments at both Facebook and
LinkedIn~\cite{copysets}, HDFS provides a reasonable estimate of distributed
filesystem performance.  Although we cannot expect WTF to grossly outperform
HDFS---both systems are limited by the speed of the hard disks in our
cluster---we can use the degree to which WTF and HDFS differ in performance to
estimate the overheads present in WTF's design.

\myparagraph{Setup} The workload for these benchmarks is generated by twelve
distinct clients, one per storage server in the cluster, that all work in
parallel.  This configuration was chosen after experimentation because
additional clients do not significantly increase the throughput, but do increase
the latency significantly.

All benchmarks operate on \unit{100}{\giga\byte} of data, or over
\unit{16}{\giga\byte} per machine once replication is accounted for.  This
workload is small enough that we can run the experiments several times each, but
is big enough to be blocked by disk on modern Linux kernels.  The Linux virtual
memory subsystem will not allow a writing process to populate the entirety of
RAM with dirty buffers; instead, only a fraction of memory may be used for dirty
pages before the kernel forces writing processes to yield time for writing back
I/O~\cite{linux-vm}.  Consequently, although our test data is not multiple times
the memory available in our cluster, it is more than five times the space
available for storing dirty buffers.  To mitigate any confounding effects of the
kernel's buffer cache on read-oriented experiments, the buffer cache was
completely cleared before each such experiment.

\begin{figure}[t]
\centering
\input{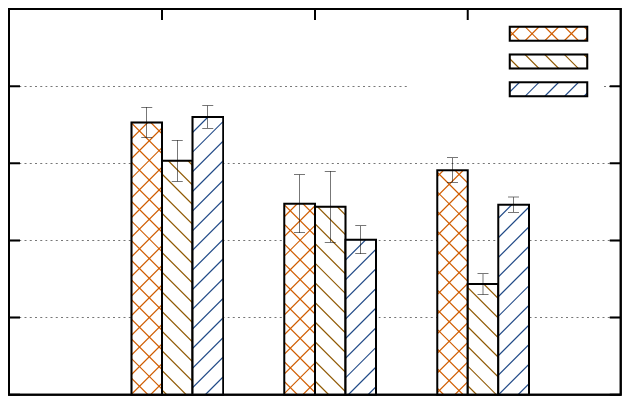}
\caption{Performance of a one-server deployment of HDFS and WTF compared with
the ext4 filesystem.  Error bars indicate the standard error of the mean across
seven trials.}
\label{fig:micro:baseline}
\vspace{-.5\baselineskip}
\end{figure}

\myparagraph{Single server performance}  This first benchmark executes on a
single server to establish the baseline performance of a one node cluster.
Here, we'll not only compare the two systems to each other, but to the same
workload implemented on a local ext4 filesystem.  Our expectation here is that
the POSIX API will provide an upper bound on performance.  To reduce the extent
to which round trip time dominates the calls in each distributed system the
client and storage server are collocated.  Figure~\ref{fig:micro:baseline} shows
the throughput of write and read operations in the one-node cluster.  From this
we can see that the maximum measured throughput of a single node in our cluster
is \unit{87}{\mega\byte\per\second}, which means the total throughput of the
cluster, assuming optimal usage, will peak at
\unit{1044}{\mega\byte\per\second}.

\begin{figure}[t]
\centering
\input{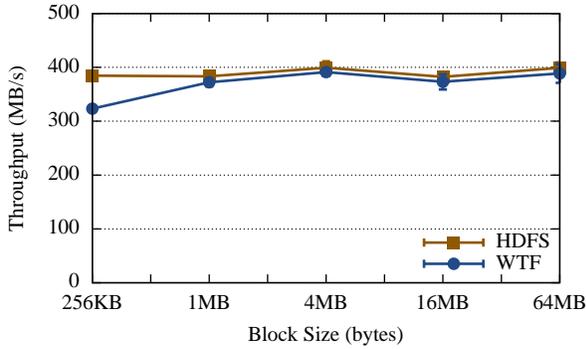}
\caption{Aggregate throughput of a sequential write workload where writers make
    fixed size calls to ``write''.  HDFS and WTF both provide applications with
    approximately \unit{400}{\mega\byte\per\second} of goodput.  Error bars
    report the standard error of the mean across seven trials.}
\label{fig:micro:write-sequential}
\vspace{-.5\baselineskip}
\end{figure}

\myparagraph{Sequential Writes}  WTF guarantees that all readers in the
filesystem see a write upon its completion; however, this guarantee is only
useful to applications when throughput remains high for smaller writes.  This
benchmark examines the impact that write size has on the aggregate throughput
achievable for filesystem-based applications by varying the block size and
measuring the aggregate throughput across all twelve writers.
Figure~\ref{fig:micro:write-sequential} shows the results for block sizes
between \unit{256}{\kilo\byte} and \unit{64}{\mega\byte}.  For writes greater
than \unit{1}{\mega\byte}, WTF achieves 97\% the throughput of HDFS.  For
\unit{256}{\kilo\byte} writes, WTF achieves 84\% of the throughput of HDFS.

\begin{figure}[t]
\centering
\input{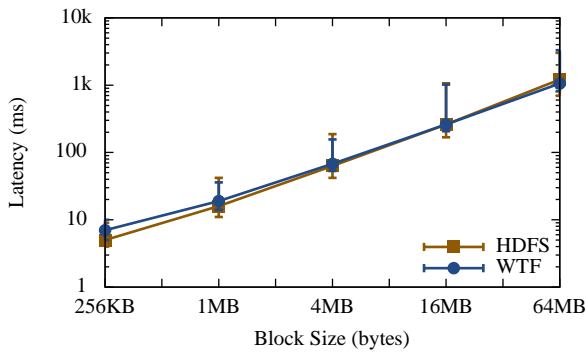}
\caption{Median latency of write operations across a variety of write sizes.
Error bars report the 5th and 95th percentile latencies.}
\label{fig:micro:write-sequential-latency}
\vspace{-\baselineskip}
\end{figure}

The latency for the two systems is similar, and directly correlated with the
block size.  Figure~\ref{fig:micro:write-sequential-latency} shows the latency
of writes across a variety of block sizes.  We can see that WTF's median latency
is very close to HDFS's median latency for larger writes, and that the 95th
percentile latency for WTF is often lower than on HDFS operations.  Latency of
WTF write operations diverges from HDFS for \unit{256}{\kilo\byte} writes.  Each
HyperDex transaction in WTF imposes an approximately \unit{3}{\milli\second}
lower bound on the total write completion time.  For the \unit{256}{\kilo\byte}
test case, this is 50\% of the median latency.  Even so, WTF's median and 95th
percentile latency measurements for this block size are only
\unit{2}{\milli\second} higher than the corresponding measurements for HDFS.

\myparagraph{Random Writes}  WTF enables applications to write at random offsets
in a file without restriction.  Because HDFS cannot support applications that
write at random offsets within a file, we cannot use it as a baseline for these
experiments; instead, the sequential write performance of WTF will serve as a
baseline to compare against the random write performance.  This this benchmark
issues writes at uniformly random offsets instead of sequentially increasing
offsets.

\begin{figure}[t]
\centering
\input{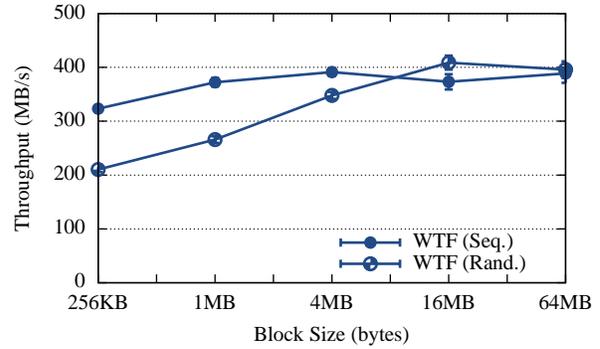}
\caption{Aggregate throughput of concurrent writers making fixed size calls to
    ``write'' at random offsets within a file.  Error bars report the standard
error of the mean across seven trials.}
\label{fig:micro:write-random}
\vspace{-.5\baselineskip}
\end{figure}

Figure~\ref{fig:micro:write-random} shows the aggregate throughput achieved by
clients randomly writing to WTF.  We can see that the random write throughput is
always within a factor of two of the sequential throughput, and that this
difference diminishes as the size of the writes approaches \unit{8}{\mega\byte}.

\begin{figure}[t]
\centering
\input{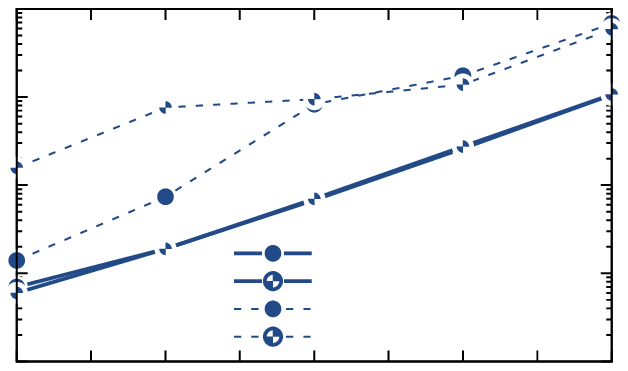}
\caption{Median and 99th percentile latencies for sequential and random WTF
    writes.  The median latency does not change between sequential and random
write patterns.}
\label{fig:micro:write-random-latency}
\vspace{-.5\baselineskip}
\end{figure}

Because the common case for a sequential write and a random write in WTF differ
only at the stage where metadata is written to HyperDex, we expect that such a
difference in throughput is directly attributable to the metadata stage.
HyperDex provides lower latency variance to applications with a small working
set than applications with a large working set with no locality of access.  We
can see the difference this makes in the tail latency of WTF writes in
Figure~\ref{fig:micro:write-random-latency}, which shows the median and 99th
percentile latencies for both the sequential and random workloads.  The median
latency for both workloads is the same for all block sizes.  For block sizes
\unit{4}{\mega\byte} and larger, the 99th percentile latencies are approximately
the same as well.  Writes less than \unit{4}{\mega\byte} in size exhibit a
significant difference in 99th percentile latency between the sequential and
random workloads.  These smaller writes spend more time updating HyperDex than
writing to storage servers.  We expect that further optimization of HyperDex
would close the gap between sequential and random write performance.

Although the difference between sequential and random performance is
significant, it is important to remember that HDFS applications cannot perform
random writes at all.  With HDFS, applications that need to change a file must
rewrite the file in its entirety, which is a costly and slow process.

\myparagraph{Sequential Reads}  Batch processing applications often read large
input files sequentially during both the map and reduce phases.  Although a
properly-written application will double-buffer to avoid small reads, the
filesystem should not rely on such behavior to enable high throughput.  This
experiment shows the extent to which WTF can be used by batch applications by
reading through a file sequentially using a fixed-size buffer.

\begin{figure}[t]
\centering
\input{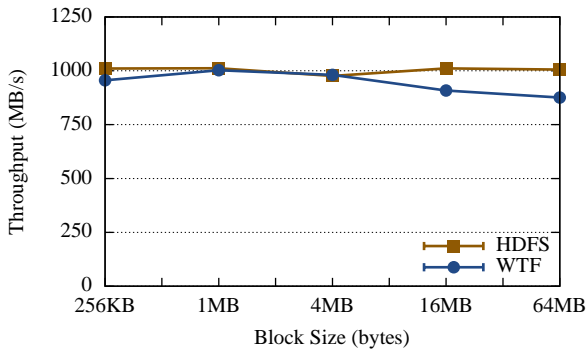}
\caption{Aggregate throughput of concurrent readers reading fixed size blocks.
    HDFS and WTF both achieve approximately \unit{900}{\mega\byte\per\second} of
    read throughput.  Error bars report the standard error of the mean across
seven trials.}
\label{fig:micro:read-sequential}
\vspace{-.5\baselineskip}
\end{figure}

Figure~\ref{fig:micro:read-sequential} shows the aggregate throughput of
concurrent readers reading through a file written by the previously described
sequential write benchmark.  We can see that for all read sizes, WTF's
throughput is at least 80\% the throughput of HDFS.  The throughput reported
here is not comparable to the throughput reported in the write benchmark because
only one of the two active replicas is consulted on each read, thus doubling the
number of disks available for independent operations.  For smaller reads, WTF's
throughput matches that of HDFS.  The difference at larger sizes is largely an
artifact of the implementations.  HDFS uses readahead on both the clients and
storage servers in order to improve throughput for streaming workloads.  By
default and in the experiment, the HDFS readahead is configured to be
\unit{4}{\mega\byte}, which is the point at which the systems start to exhibit
different characteristics.  Our preliminary WTF implementation does not have any
readahead mechanism, and exhibits higher latency.  A more mature implementation
could take advantage of readahead to reduce this difference.

\myparagraph{Random Reads}  Applications built on a distributed filesystem, such
as key-value stores or record-oriented applications often require random access
to the files.  This experiment shows the performance of applications reading
constant-sized pieces from a file at offsets that are chosen uniformly at
random.

\begin{figure}[t]
\centering
\input{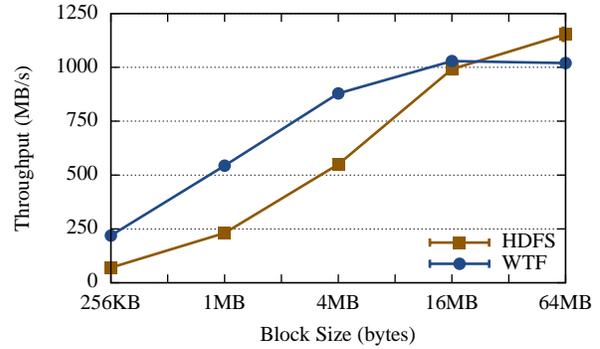}
\caption{Aggregate throughput of random reads of varying size in a
    two-replicated deployment.  We can see that WTF-backed applications achieve
    higher throughput than HDFS applications for a variety of small read sizes.
Error bars indicate the standard error of the mean across seven trials.}
\label{fig:micro:read-random}
\vspace{-\baselineskip}
\end{figure}

Figure~\ref{fig:micro:read-random} shows the aggregate throughput of twelve
concurrent random readers.  We can see that for reads of less than
\unit{16}{\mega\byte}, WTF achieves significantly higher throughput---at its
peak, WTF's throughput is $2.4\times$ the throughput of HDFS.  Here, the
readahead and client-side caching that helps HDFS with larger sequential read
workloads adds overhead to HDFS that WTF does not incur.  The 
95th percentile latency of a WTF read is
less than the median latency of a HDFS read for block sizes less than
\unit{4}{\mega\byte}.

\myparagraph{Scaling the Workload} This experiment varies the number of clients
writing to the filesystem to explore how concurrency
affects both latency and throughput.  This benchmark employs the workload from
the sequential-write benchmark with a \unit{4}{\mega\byte} write
size and a variable number of workload-generating clients.

\begin{figure}[t]
\centering
\input{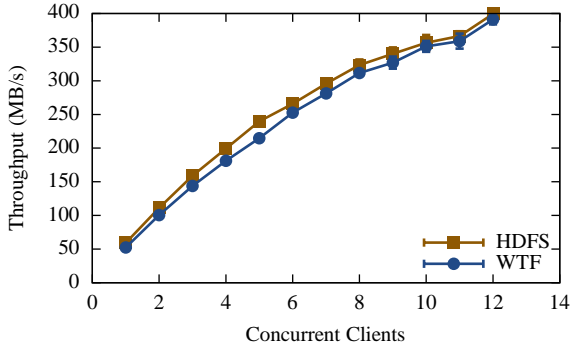}
\caption{Aggregate throughput as the number of writers increases.  Error bars
show the standard error of the mean across seven trials.}
\label{fig:micro:write-sequential-minions}
\vspace{-.5\baselineskip}
\end{figure}

\begin{figure}[t]
\centering
\input{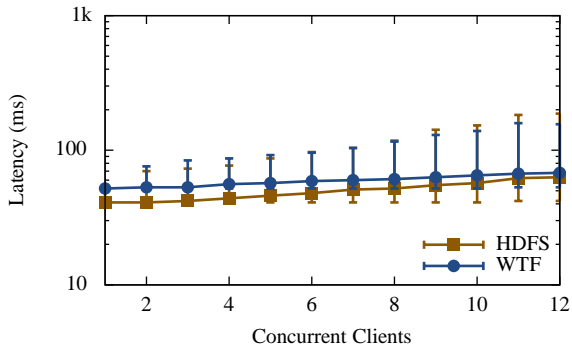}
\caption{Median write latency as the number of writers increases.  Error bars
show the 5th and 95th percentile latencies.}
\label{fig:micro:write-sequential-minions-latency}
\vspace{-\baselineskip}
\end{figure}

Figure~\ref{fig:micro:write-sequential-minions} shows the resulting throughput
for between one and twelve clients.  We can see that the single client
performance is approximately \unit{60}{\mega\byte\per\second}, while twelve
clients sustain an aggregate throughput of approximately
\unit{380}{\mega\byte\per\second}.  WTF's throughput is approximately the same
as the throughput of HDFS for each data point.  Running the same workload with
forty-eight clients did not increase the throughput beyond the throughput
achieved with twelve clients.  We can see the corresponding latency change in
Figure~\ref{fig:micro:write-sequential-minions-latency}.

\myparagraph{Garbage Collection}

\begin{figure}[t]
\centering
\input{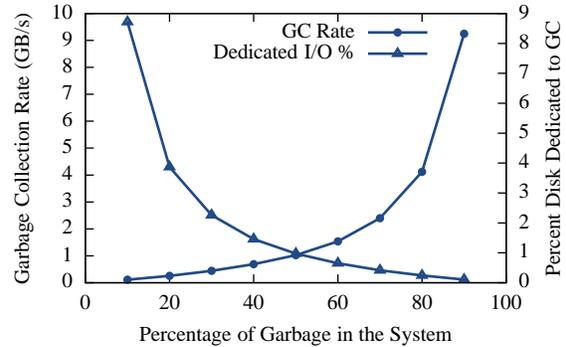}
\caption{The maximum rate of garbage collection is positively correlated with
    the amount of garbage to be collected.  Consequently, WTF dedicates a small
fraction of its overall I/O to garbage collection.}
\label{fig:micro:gc}
\vspace{-\baselineskip}
\end{figure}

This benchmark measures the overhead of garbage collection on a storage server.
As mentioned in Section~\ref{sec:gc}, it is more efficient to collect files with
more garbage than files with less garbage, and WTF preferentially garbage
collects these larger files.  Figure~\ref{fig:micro:gc} shows the rate at which
the cluster can collect garbage, for varying amounts of randomly located
garbage, when all resources are dedicated to the task.  We can see that when the
cluster consists of 90\% garbage, the cluster can reclaim this garbage at a rate
of over \unit{9}{\giga\byte} of garbage per second, because it need only write
\unit{1}{\giga\byte\per\second} to reclaim the garbage.

It is, however, impractical to dedicate all resources to garbage collection;
instead, WTF dedicates only a fraction of I/O to the task.  Storage servers
initiate garbage collection when disk usage exceeds a configurable threshold,
and ceases when the amount of garbage drops below 20\%.
Figure~\ref{fig:micro:gc} shows that the maximum overhead required to maintain
the system below this threshold is 4\%.

\section{Related Work}

Filesystems have been an active research topic since the earliest days of
systems research.  Existing approaches related to WTF can be broadly classified
into two categories based upon their design.

\myparagraph{Distributed filesystems}

Distributed filesystems expose one or more units of storage over a network to
clients.  AFS~\cite{afs} exports a uniform namespace to workstations, and stores
all data on centralized servers.   Other systems~\cite{zebra,gpfs,lustre}, most
notably xFS~\cite{xfs} and Swift~\cite{swift} stripe data across multiple
servers for higher performance than can be achieved with a single disk.
Petal~\cite{petal} provides a virtual disk abstraction that clients may use as a
traditional block device.  Frangipani~\cite{frangipani} builds a filesystem
abstraction on top of Petal.  NASD~\cite{nasd} and Panasas~\cite{panasas} employ
customized storage devices that attach to the network to store the bulk of the
metadata.  In contrast to these systems, WTF provides transactional guarantees
that can span hundreds or thousands of disks because its metadata storage scales
independently of the number of storage servers.

Recent work has focused on building large-scale datacenter-centric filesystems.
GFS~\cite{gfs} and HDFS~\cite{hdfs} employ a centralized master server that
maintains the metadata, mediates client access, and coordinates the storage
servers.  Salus~\cite{salus} improves HDFS to support storage and computation
failures without loss of data, but retains the central metadata server.  This
centralized master approach, however, suffers from scalability bottlenecks
inherent to the limits of a single server~\cite{gfsacmq}.  WTF overcomes the
metadata scalability bottleneck using the scalable HyperDex key-value
store~\cite{warp}.

CalvinFS~\cite{calvinfs} focuses on fast metadata management using distributed
transactions in the Calvin~\cite{calvin} transaction processing system.
Transactions in CalvinFS are limited, and cannot do read-modify-write operations
on the filesystem without additional mechanism.  Further, CalvinFS addresses
file fragmentation using a heavy-weight garbage collection mechanism that
entirely rewrites fragmented files; in the worst case, a sequential writer could
incur I/O that scales quadratically in the size of the file.  In contrast, WTF
provides fully general transactions and carefully arranges data to improve
sequential write performance.

Another approach to scalability is demonstrated by Flat Datacenter
Storage~\cite{fds}, which enables applications to access any disk in a cluster
via a CLOS network with full bisection bandwidth.  To eliminate the scalability
bottlenecks inherent to a single master design, FDS stores metadata on its tract
servers and uses a centralized master solely to maintain the list of servers in
the system.  Blizzard~\cite{blizzard} builds block storage, visible to
applications as a standard block device, on top of FDS, using nested striping
and eventual durability to service the smaller writes typical of POSIX
applications.  These systems are complementary to WTF, and could implement the
storage servers abstraction.

Power-proportional filesystems are elastic, in that they dynamically change the
power consumption of a cluster to scale resource usage with demand and decrease
power consumption in the cluster~\cite{rabbit,sierra,springfs}.  WTF's design
does not consider power-proportionality, but could possibly incorporate
allocation techniques from other systems to make it more elastic.

Other ``blob'' storage systems behave similarly to file systems, but with a
restricted interface that permits creating, retrieving, and deleting blobs,
without efficient support for arbitrarily changing or resizing blobs.
Facebook's f4~\cite{f4} ensures infrequently accessed files are readily
available for access.  Pelican~\cite{pelican} enables power-efficient cold
storage by over provisioning storage space, and selectively turning on subsets
of the disks to service requests.  The design goals of these systems are
different from the interactive, online applications that WTF enables; WTF could
be used in front of these systems to generate, maintain, and modify data before
placing it in warm or cold storage.

\myparagraph{Transactional filesystems}

Transactional filesystems enable applications to offload much of the hard work
relating to update consistency and durability to the filesystem.  The
QuickSilver operating system shows that transactions across the filesystem
simplify application development~\cite{quicksilver}.  Further work showed that
transactions could be easily added to LFS, exploiting properties of the
already-log-structured data to simplify the design~\cite{xactlfs}.
Valor~\cite{valor} builds transaction support into the Linux kernel by
interposing a lock manager between the kernel's VFS calls and existing VFS
implementations.  In contrast to the transactions provided by WTF, and the
underlying HyperDex transactions, these systems adopt traditional pessimistic
locking techniques that hinder concurrency.

Optimistic concurrency control schemes often enable more concurrency for
lightly-contended workloads.  PerDiS FS adopts an optimistic concurrency control
scheme that relies upon external components to reconcile concurrent changes to a
file~\cite{perdis}.  This allows users and applications to concurrently work on
the same file; according to the authors, the most commonly adopted technique is
selecting one version and throwing the rest away.  Liskov and Rodrigues show
that much of the overhead of a serializable filesystem can be avoided by running
read-only transactions in the recent past, and employing an optimistic protocol
for read-write transactions~\cite{liskovfast}.  WTF builds on top of HyperDex's
optimistic concurrency and supports operations such as \code{append} that avoid
creating conflicts between concurrent transactions.

WTF is not the first system to choose to employ a transactional database as part
of its design.  Inversion~\cite{inversion} builds on PostgreSQL to maintain a
complete filesystem.  KBDBFS~\cite{valor} and Amino~\cite{amino} both build on
top of BerkeleyDB; the former is an in-kernel implementation of BerkeleyDB,
while the latter eschews the complexity and takes a performance hit with a
userspace implementation.  WTF differs from these designs in that it stores
solely the metadata in the transactional data store; data is stored elsewhere
and not managed by the transactional component.  Further, its design ensures
that transactions on metadata are sufficient to provide filesystem-level
transactions.

Stasis~\cite{stasis} makes the argument that no one design support all use
cases, and that transactional components should be building blocks for
applications.  WTF's approach is similar:  HyperDex's transactions are used as a
base primitive for managing WTF's state, and WTF supports a transactional API.
Applications built on WTF can use this API to achieve their own transactional
behavior.

\section{Conclusion}

This paper described the Wave Transactional Filesystem (WTF), a new distributed
filesystem that enables applications to operate on multiple files
transactionally without requiring complex application logic.  A new filesystem
abstraction called {\em file slicing} further boosts performance by modifying
files more efficiently than traditional primitives permit.  The main insight
behind file slicing is that it enables applications to read and write using
references to data that is stored elsewhere in the filesystem.

A broad evaluation shows that WTF achieves throughput and latency similar to
industry-standard HDFS, while simultaneously offering stronger guarantees and a
richer API.  A sample application built with file slicing outperforms
traditional approaches by a factor of four by reducing the overall I/O cost.

The ability to make transactional changes to multiple files at scale is novel in
the distributed systems space, and the file slicing APIs enable a new class of
applications that are difficult to implement efficiently with current APIs.
Together, these features are a potent combination that enables a new class of
high performance applications.

\hbadness=10000
\bibliographystyle{plain}
\bibliography{wtf}

\end{document}